\documentclass{aa}
\newcommand{\bit}{\begin{itemize}}
\newcommand{\eit}{\end{itemize}}
\newcommand{\ben}{\begin{enumerate}}
\newcommand{\een}{\end{enumerate}}
\newcommand{\bde}{\begin{description}}
\newcommand{\ede}{\end{description}}

\usepackage{graphicx}
\usepackage{lscape}

\begin{document}
\title{2D velocity fields of simulated interacting disc galaxies}
\author{T. Kronberger$^{1,2}$,
        W. Kapferer$^1$,
        S. Schindler$^1$,
        and B. L. Ziegler$^{2,3}$}

\institute{ $^1$Institut f\"ur Astro- und Teilchenphysik,
            Universit\"at Innsbruck,
            Technikerstr. 25,
            A-6020 Innsbruck, Austria \\
            $^2$Institut f\"ur Astrophysik,
            Universit\"at G\"ottingen,
            Friedrich-Hund-Platz 1,
            D-37077 G\"ottingen, Germany \\
            $^3$Argelander-Institut f\"ur Astronomie,
            Universit\"at Bonn,
            Auf dem H\"ugel 71,
            D-53121 Bonn, Germany}

\offprints{T. Kronberger, \email{Thomas.Kronberger@uibk.ac.at}}

\date{-/-}

\abstract{We investigate distortions in the velocity fields of disc
galaxies and their use to reveal the dynamical state of interacting
galaxies at different redshift. For that purpose, we model disc
galaxies in combined N-body/hydrodynamic simulations. 2D velocity
fields of the gas are extracted from these simulations which we
place at different redshifts from z$=$0 to z$=$1 to investigate
resolution effects on the properties of the velocity field. To
quantify the structure of the velocity field we also perform a
kinemetry analysis. If the galaxy is undisturbed we find that the
rotation curve extracted from the 2D field agrees well with
long-slit rotation curves. This is not true for interacting systems,
as the kinematic axis is not well defined and does in general not
coincide with the photometric axis of the system. For large (Milky
way type) galaxies we find that distortions are still visible at
intermediate redshifts but partly smeared out. Thus a careful
analysis of the velocity field is necessary before using it for a
Tully-Fisher study. For small galaxies (disc scale length $\sim$ 2
kpc) even strong distortions are not visible in the velocity field
at $z\approx0.5$ with currently available angular resolution.
Therefore we conclude that current distant Tully-Fisher studies
cannot give reliable results for low-mass systems. Additionally to
these studies we confirm the power of near-infrared integral field
spectrometers in combination with adaptive optics (such as SINFONI)
to study velocity fields of galaxies at high redshift (z$\sim$2).

}

\authorrunning {T. Kronberger et al.}
\titlerunning {2D velocity fields of simulated interacting disc galaxies}

\maketitle
%

\section{Introduction}
Recently it has become technically feasible to observe the full 2D
velocity field (VF) of local galaxies in optical wavebands using
integral field units (IFUs) such as SAURON (e.g. Ganda et al. 2006)
or Fabry-Perot interferometry (e.g. Chemin et al. 2006, Garrido et
al. \cite{Garrido}). For intermediate and high redshift galaxies,
however, there are by now hardly any observational studies of 2D
velocity fields available.

Flores et al. (2006) observed the 2D velocity field of 35 galaxies
at intermediate redshift (0.4 $<$ z $<$ 0.75) using FLAMES/GIRAFFE
at VLT. One aspect was to investigate the redshift evolution of the
Tully-Fisher relation. A different approach, used by our group, was
presented by Ziegler et al. (2006) and Kutdemir et al. (2007) who
utilize multiple-object spectroscopy from the VLT with different
slit positions on each galaxy in order to construct the full
velocity field for each galaxy. Most of the other studies of
distant, faint, and small galaxies are still based on slit
spectroscopy (e.g. Vogt 2001, B\"ohm et al. 2004). To account for
distortions and irregularities in the velocity fields is in both
cases critical, especially when aiming at a distant Tully-Fisher
study. In two recent papers we showed that observational constraints
and galaxy--galaxy interactions can severely influence the
determination of the rotation curve of observed disc galaxies
(Kapferer et al. 2006, Kronberger et al. 2006).

In this paper we investigate to what extent the full 2D velocity
field of a galaxy can be used to gain information on its internal
kinematics and can hence improve the quality of e.g. Tully-Fisher
studies. This question is also important to possibly disentangle
different interaction processes by mapping 2D velocity fields and to
study their impact on galaxy evolution. We focus on the question how
the visibility of distortions depends on the redshift of the
observed galaxy, i.e. the actual spatial resolution of the galaxy.
For that investigation we place a galaxy at different redshifts and
bin the velocity according to the spatial resolution at this
redshift. A similar study for observed galaxies was presented by
Epinat et al. (2006).

Additionally to the examination of 2D velocity fields of
intermediate redshift galaxies from optical spectroscopy we
investigate the performance of near-infrared integral field
spectrographs that are used together with adaptive optics. As a
prototype we take the characteristics of SINFONI at the Very Large
Telescope of the European Southern Observatory to study the velocity
fields of galaxies at z$\sim$2. For example, Genzel et al. (2006)
observed the velocity field of a massive protodisc at z=2.38
detecting an ordered rotation without any hint for a major merger
event in the system.

Very recently also Jesseit et al. (2007) analysed 2D velocity fields
of simulated galaxies. They focused on simulated disc merger
remnants and found that many different kinematical phenomena can be
observed in the stellar velocity maps, such as kinematic misaligned
discs or counter-rotating-cores. For the analysis they also used the
kinemetric method of Krajnovi\'{c} et al. (2006) as we do in this
paper. They did, however, not investigate the redshift dependence of
the 2D velocity fields.

The paper is organised as follows. In Sect. 2 and 3, we describe the
simulations, the interaction geometries, and the way we extract
realistic 2D velocity fields from the numerical data. In Sect. 4 the
results for different interaction scenarios and their dependence on
the angular resolution are presented. We end with a summary of the
main conclusions in Sect. 5.

\section{Simulations}
In this work we use some of the simulated systems presented in
Kapferer et al. (2005), which were subsequently used by Kapferer et
al. (2006) and Kronberger et al. (2006). The simulations were
carried out with the N-body/SPH code GADGET-2 developed by V.
Springel (see Springel 2005 for details). In this code the gas of
the galaxies is treated hydrodynamically and prescriptions for
cooling, star formation, stellar feedback, and galactic winds are
included (Springel \& Hernquist, 2003). The collisionless dynamics
of the dark matter and the stellar component is calculated using an
N-body technique. The initial conditions were built according to
Springel et al. (2005), based on the analytic work of Mo et al.
(1998). The two model galaxies were chosen such that they represent
a Milky Way type and a small spiral galaxy, with the mass ratio of
the two galaxies being 8:1. Therefore, the total mass of the model
galaxies A and B is 1.34$\times$ 10$^{12}$ $h^{-1}$ $M_{\sun}$ and
1.67188$\times$10$^{11}$ $h^{-1}$ $M_{\sun}$, respectively. The
combined N-body/SPH simulation then calculates 5 Gyr of evolution.
For every time step, we know the velocity of each particle and can
hence extract realistic 2D velocity fields of the gas.

Concerning the spatial alignment and impact parameters, we follow
the notation introduced by Duc et al. (2000) to describe the
interaction geometry. The parameter {\bf b} corresponds to the
minimum separation of the galaxies' trajectories, as if they were
point masses on Keplerian orbits. Additionally, two angles, $\Theta$
and $\Phi$ define the spatial orientation of the disc.

For the complete sample, we selected the alignments in such a way as
to cover as many geometries as possible, including minor and major
mergers and fly-bys (achieved by increasing the minimum separation).
For this analysis we just use three different interaction
geometries, which we list in Table \ref{sims}. These three
simulations allow us to study different classes of kinematical
distortions, detailed in Sect. \ref{extract}.

\begin{table*}
\caption[]{Interaction parameters for the simulations. C2: the two
interacting model galaxies. Model galaxy A is a Milky Way type
spiral galaxy while model galaxy B represents a small spiral galaxy,
with the mass ratio of the two galaxies being 8:1.; C7: minimum
separation [kpc]; C8: initial relative velocities of the two
galaxies [km/s].In column C9 we give the initial orbital energy of
the encounter in Joule.}
\begin{center}
\begin{tabular}{c c c c c c c c c}
\hline \hline Simulation & C2 & $\Phi_1$ & $\Theta_1$ & $\Phi_2$ &
$\Theta_2$ & C7 & C8 & C9 \cr \hline 1 & A-B & 0 & 0 & 0 & 0 & 50 &
250 & $4.8 \times 10^{51}$\cr 2 & A-B & 0 & 0 & 0 & 0 & 5 & 250
&$4.8 \times 10^{51}$ \cr 3 & B-B & 0 & 0 & 0 & 0 & 5 & 120& $9.7
\times 10^{50}$ \cr 4 & A-B & 0 & 0 & 180 & 0 & 5 & 250 & $4.8
\times 10^{51}$ \cr 5 & A-B & 0 & 0 & 90 & 0 & 5 & 250 & $4.8 \times
10^{51}$ \cr 6 & A-A & 0 & 0 & 0 & 0 & 5 & 350 & $1.1 \times
10^{53}$\cr \hline
\end{tabular}
\label{sims}
\end{center}
\end{table*}

Additional important quantities of the simulations, as e.g. the
particle numbers, are summarized in Table
\ref{galaxyproperties_resolution}. Throughout the paper we adopt the
standard $\Lambda$CDM cosmology with $\Omega_\Lambda=0.7$,
$\Omega_m=0.3$, and h=0.7.

\begin{table*}
\begin{center}
\caption[]{Particle numbers, mass resolution and gravitational
softening used for the two model galaxies. Additionally the circular
velocity of the halo at $\rm{r}_{200}$ and the disc scale length of
the initial conditions are given.}
\begin{tabular}{c c c c c c}
\hline \hline & particle number & mass resolution & softening length
& halo circular velocity & radial disc \cr& & [$h^{-1}$
$M_{\sun}$/particle] & [$h^{-1}$ kpc] & [km/s] & scale length
[$h^{-1}$ kpc]\cr\hline Galaxy A:& & & & &\cr \hline Dark matter
halo & 30000 &$4.2\times10^{6}$ & 0.4 & 160 &\cr Disk collisionless
& 20000 & $3.3\times10^{5}$& 0.1 & & \cr Gas in disk & 35000 &
$8.4\times10^{4}$& 0.1& &4.5\cr \hline Galaxy B:& & & \cr \hline
Dark matter halo & 30000 & $5.3\times10^{5}$& 0.4& 80& \cr Disk
collisionless & 20000 & $4.2\times10^{4}$& 0.1& & \cr Gas in disk &
35000 & $1.1\times10^{4}$& 0.1 & &2.25\cr
\end{tabular}
\label{galaxyproperties_resolution}
\end{center}
\end{table*}

\section{Extraction of realistic 2D velocity fields} \label{extract}

In order to construct realistic 2D velocity fields, we project all
gas particles of the N-body/SPH simulations onto a Cartesian,
equally spaced grid. The spacing is chosen such, that the spatial
resolution at the assumed redshift of the model galaxy corresponds
to the angular resolution of current state-of-the-art observations.
For the whole investigated redshift range from z=0 to z=1.0 we adopt
an angular resolution typical for IFU or FPI observations, namely
0.4" (e.g. Chemin et al. 2006). The angular resolution of SAURON,
for example, would be 0.3" or 0.9". For intermediate redshifts we
additionally chose 0.25", which is the pixel scale along the FORS2
slit as used by Ziegler et al. (2006) and 0.52", which is the
lenslet size of FLAMES/GIRAFFE at VLT (e.g. Flores et al. 2006). We
calculate for each redshift the physical resolution according to the
given angular resolution using the concordance cosmological model
(see Table \ref{redshift}). The velocity field of the galaxy is
binned using this spatial resolution.

\begin{table}
\caption[]{Physical resolution according to 1 arcsec angular
resolution for different redshifts in the standard $\Lambda$CDM
cosmology.}
\begin{center}
\begin{tabular}{c c}
\hline \hline Redshift & physical resolution according to 1" [kpc]
\cr \hline 0.05 & 0.98 \cr 0.1 & 1.85 \cr 0.2 & 3.30 \cr 0.3 & 4.46
\cr 0.4 & 5.38 \cr 0.5 & 6.11 \cr 0.8 & 7.51 \cr 1.0 & 8.01 \cr 2.0
& 8.38 \cr \hline
\end{tabular}
\label{redshift}
\end{center}
\end{table}

From the knowledge of the full 3D velocity fields and the
interaction history of the galaxies in our simulations we define
three kinematical classes, in principle analogue to classifications
in Flores et al. (2006) or Krajnovi\'{c} et al. (2006):
\begin{itemize}
\item \textbf{Undisturbed Rotation:} The galaxy is not interacting
and the rotation curve has the "classical" shape, rising in the
inner part and turning over to a flat regime.
\item \textbf{Disturbed Rotation:} Rotation is still
predominant but disturbed by a minor interaction, e.g. a minor
merger event. Non-rotational components of the velocity field are
also clearly visible as peculiarities in 1D rotation curves
(Kronberger et al. 2006).
\item \textbf{Distorted Velocity Field:} The velocity field is
heavily distorted by a strong merger event, i.e. an interaction with
a galaxy of similar or higher mass. It shows a very complex pattern
and has no regular rotation curve anymore.
\end{itemize}

The main question that we investigate in this paper is how such
classifications depend on the redshift of the observed galaxy, i.e.
the actual spatial resolution of the galaxy. In this context the
seeing plays a crucial role, as it typically exceeds the angular
resolution of the instrument. To simulate seeing effects on our
velocity-field measurements, a convolution with a Gaussian point
spread function was applied. We adopted a value of 0.8" for the FWHM
of the Gaussian seeing, which is a typical value for ground based
observations (see e.g. J\"ager et al. 2004). The appearance of the
velocity fields is therefore dominated by the seeing. Note that we
do not calculate an evolution of properties of the galaxy with
redshift but study how a given kinematical state of a galaxy is
observed at different redshifts.

All studies mentioned above are mainly seeing dominated. We
additionally study the possibility to identify mergers at high
redshift (z$\sim$2) with an adaptive optics instrument such as
SINFONI at the VLT. For this investigation we adopt an angular
resolution of 0.15" as achieved by Genzel et al. (2006).

\section{Results}

In the subsequent sub-sections we will systematically investigate
velocity fields from each of the kinematical classes defined in
Sect. \ref{extract}. The focus lies on the visibility of kinematical
distortions as a function of redshift. In order to quantify the
distortions and to interpret the partly complex structures in the
velocity fields we use the kinemetry package of Krajnovi\'{c} et al.
(2006). This analysis is based on the assumption, that the mean
velocity along best fitting ellipses can be reproduced by a cosine
law, i.e.

\begin{equation}
V(a,\Psi)=V_0+V_c(a)cos{\Psi},
\end{equation}

\noindent where $a$ is the length of semi-major axis of the ellipse,
$\Psi$ is for discs the azimuthal angle measured from the major axis
in the plane of the galaxy. Note that in general a simple sine
correction for the inclination of the disc is applied in
observations too. The velocity fields presented here have not been
corrected in this way as all of them were extracted at the same
inclination i$=$35$^{\circ}$ (except for those presented in Sect.
\ref{biases}).

The position angle $\Gamma$ and the axial ratio ($q=b/a$) of the
ellipses are calculated as a function of radius from the galactic
centre. Deviations from that cosine law are measured using an
harmonic expansion along the ellipses, i.e.

\begin{equation}
V(a,\Psi)=\sum_{n=1}^{N}k_n(a)cos[n(\Psi-\phi_n(a))],
\end{equation}

\noindent where $\phi$ is the phase coefficient. Higher order
Fourier terms and radial changes of $\Gamma$ or $q$ are used to
quantize deviations in the velocity field from a simple rotation.
There is an intrinsic relation of the surface brightness of the
galaxy and the line-of-sight velocity. Both are moments of the
underlying distribution function, where the surface brightness is
the zeroth's order moment and the line-of-sight velocity is the
first order moment. Note, however, that for gas, as in our case, the
situation is different from collisionless stellar systems, where the
distribution function for simplified systems can be calculated
analytically. As we deal here mostly with rotating disc-like
structures the use of ellipses is a natural choice. For distorted
and elliptical systems this assumption will be violated and the
deviation is quantified by the Fourier analysis. In principle the
calculation of the best fitting ellipses works in a two step
approach as follows. First, higher order terms are minimized, which
do, according to their analysis, not carry the major contribution to
the map. This happens on a grid of position angles ($\Gamma$) and
flattenings ($q$) and leads to initial values ($\Gamma_{min}$) and
($q_{min}$) for a second fit. In this second fit all parameters of
the ellipse are taken into account and the Fourier analysis can be
performed to arbitrary order. For more details on the algorithm see
Krajnovi\'{c} et al. (2006).

\subsection{Undisturbed velocity fields}\label{UR}

\begin{figure}
\begin{center}
{\includegraphics[width=\columnwidth]{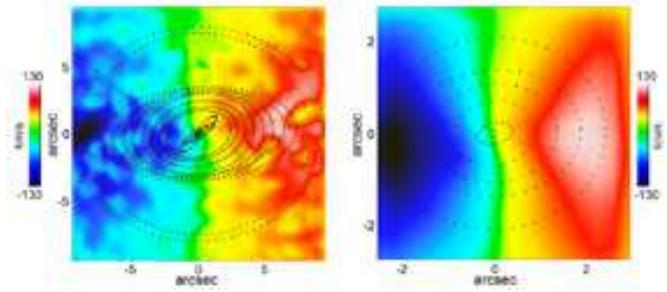}} \caption{2D
velocity field of an undisturbed modelled disc galaxy as seen at
redshift 0.1 (left) and at 0.5 (right). Overlayed as contours are
the best fitting ellipses from the kinemetry analysis.}
\label{regular0105}
\end{center}
\end{figure}

As a first step we investigate the simple case of an undisturbed
spiral galaxy, i.e. an example for the first kinematical class
presented in Sect. \ref{extract}. This class is particularly
important for Tully - Fisher studies, as only for these galaxies,
the maximum rotational velocity V$_{max}$ is an approximation for
the virial mass of the system. For this investigation we use galaxy
A in simulation 1 (see Table \ref{sims}), at a snapshot when the two
galaxies are still well separated. The adopted inclination of the
galaxy is 35$^\circ$ (90$^\circ$ is defined to be edge-on). The
velocity field is very regular at all redshifts. The appearance of
our VFs is dominated by seeing, as we adopted a constant value of
0.8" for the FWHM, which is always larger than the angular
resolution we use. Both effects, the worse sampling at higher
redshifts and the large seeing smear out the velocity field when
shifted to higher z. In Fig. \ref{regular0105} we present the VFs of
an undisturbed disc galaxy as seen at two different redshifts (z=0.1
and z=0.5). Overlayed on the VF are the best fitting ellipses from
the kinemetry analysis. At a redshift of 0.1 there is some structure
visible in the VF, e.g. a small twist of the position angle $\Gamma$
towards the centre. Also in the disc some fluctuations of the
rotational velocity are present. These small structures in the VF
are completely smeared out at redshift $z=0.5$.

\begin{figure}
\begin{center}
{\includegraphics[width=\columnwidth]{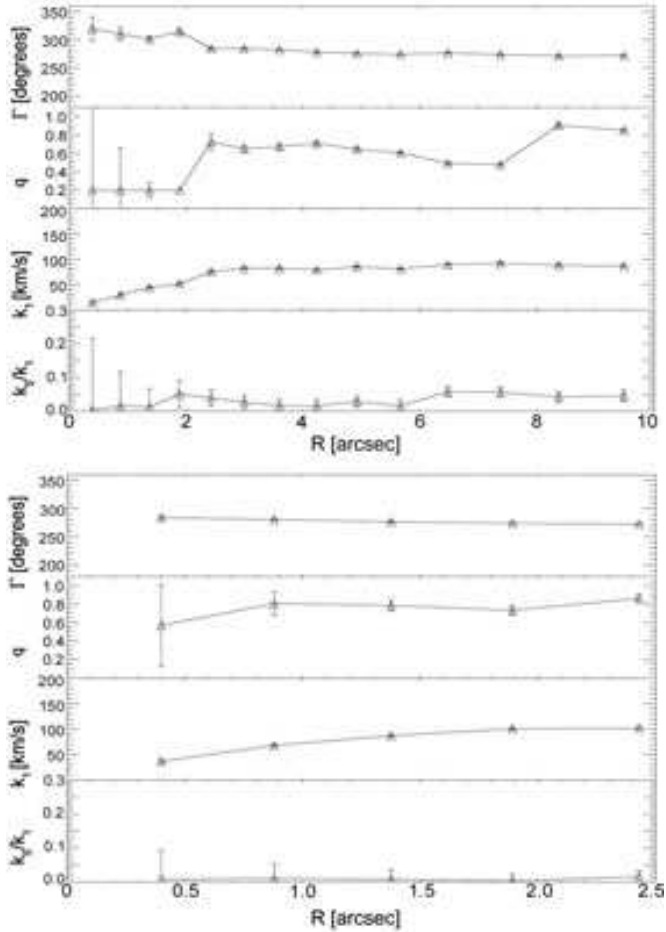}} \caption{Radial
profiles of the kinemetric properties, calculated using the
kinemetry programme for an undisturbed Milky Way type galaxy at
redshift $z=0.1$ (top) and $z=0.5$ (bottom). The position angle
$\Gamma$ and the flattening $q$ of the best fitting ellipses as well
as the first and the fifth order Fourier terms $k_1$ and $k_5$ are
plotted as a function of radius.} \label{plotsreg}
\end{center}
\end{figure}

Quantitatively the differences can be seen in Fig. \ref{plotsreg} in
the radial profiles of the kinemetric properties, calculated using
the kinemetry programme. The program also calculates formal
1$\sigma$ errors from the covariance matrix, which are shown as
error bars in the plots. These uncertainties are estimated from the
measurement uncertainties in the kinematic data. As we do not
account for the spectral resolution of the instruments we use the
scatter of the velocity field. Note that this is just a formal
uncertainty, which in reality is higher, also because systematical
errors add up.

\begin{figure}
\begin{center}
{\includegraphics[width=\columnwidth]{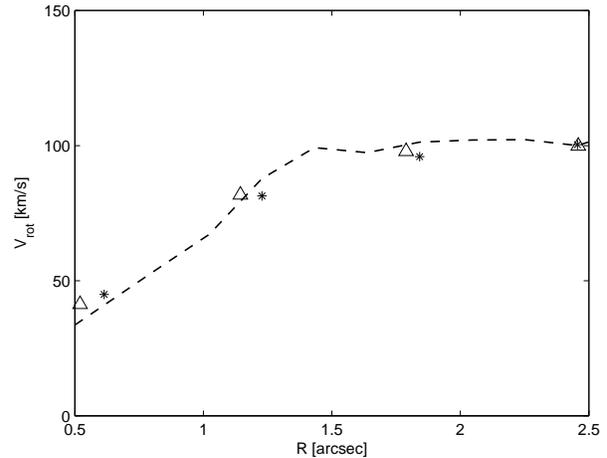}} \caption{Rotation
curves of the isolated galaxy extracted from the 2D velocity field
(triangles) and using a modelled slit (asterisks) at redshift z=0.5.
The dashed line shows the corresponding RC at redshift z=0.1
obtained from a modelled slit.} \label{RCs_regular}
\end{center}
\end{figure}

While at $z=0.1$ both, $\Gamma$ and $q$ show variations with radius,
they are almost constant at $z=0.5$. The first order moment $k_1$
corresponds to the rotation curve (RC) of a spiral galaxy or more
generally spoken to the bulk motion in the velocity field. For the
undisturbed velocity field presented here it shows the typical
behaviour of a rotation curve, i.e. rising in the inner part and
turning over to a flat regime. This undisturbed shape is present at
redshift $z=0.1$ and at $z=0.5$. In Fig. \ref{RCs_regular} we show
the rotation curves of the system at redshift $z=0.5$ obtained from
the 2D velocity field (triangles) and from a simulated slit
(asterisks), extracted as described in Kronberger et al. (2006). The
modelled slit was placed over the major axis of the galaxy with a
slit width of 1". The RCs extracted in these different ways agree
very well, as expected for such a regular velocity field. As a
dashed line we plot in the same figure the RC extracted at redshift
$z=0.1$, which agrees reasonably well with the higher redshift RCs.
This demonstrates the principle power of 2D velocity fields for
distant Tully-Fisher studies \textit{if} regular, undisturbed
velocity fields are considered.

The last row in Fig. \ref{plotsreg} shows a quantitative measure for
distortions in the 2D velocity field. The fifth order term of the
harmonic expansion $k_5$ represents complex, kinematically separate
components in the velocity field. In the plot the ratio $k_5/k_1$ is
shown. For the regular VF presented in this section, this ratio is
small, generally below 0.1 for all redshifts. However, this value is
very sensitive to the resolution of the velocity field. While at
$z=0.1$ there are some radii with increased $k_5/k_1$, these
signatures are completely smeared out at higher redshifts. At
$z=0.5$ the ratio is almost constant below a value of 0.02. This
already indicates problems with identifying distortions in 2D
velocity fields at higher redshifts. This issue will be addressed in
the next section.

At redshift $z=0.5$ we study the effects of angular resolution on VF
properties by considering in addition to the standard angular
resolution of 0.4" two more values. We adopt additionally the 0.25"
of Ziegler et al. (2006) in their sample using FORS2 spectroscopy
and 0.52" as for FLAMES/GIRAFFE at VLT (e.g. Flores et al. 2006). At
the FLAMES/GIRAFFE angular resolution the VF is even more smeared
out than at 0.4" resolution. However, the radial behaviour of the
kinemetry parameters is only slightly affected. Especially the RCs
and their maximum value V$_{max}$ agree reasonably well for all
three angular resolutions. Again, with respect to subsequent
sections, we point out that this holds only for regular velocity
fields.

\subsection{Disturbed rotation}\label{DR}

\begin{figure}
\begin{center}
{\includegraphics[width=\columnwidth]{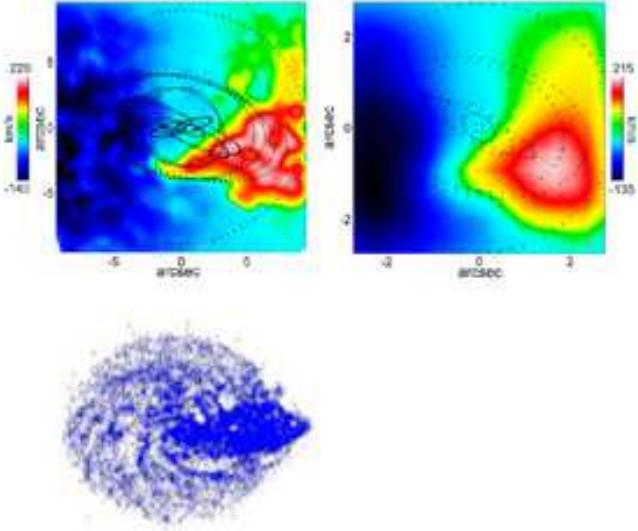}} \caption{2D
velocity field of a modelled Milky Way type disc galaxy disturbed by
an ongoing minor merger event (i.e. the small galaxy B is permeating
galaxy A in its first passage) as seen at redshift 0.1 (top left)
and 0.5 (top right). Overlayed as contours are the best fitting
ellipses from the kinemetry analysis. In the lower panel the
projected gas distribution at this snapshot is shown to illustrate
the interaction geometry.} \label{disturbed0105}
\end{center}
\end{figure}

\begin{figure}
\begin{center}
{\includegraphics[width=\columnwidth]{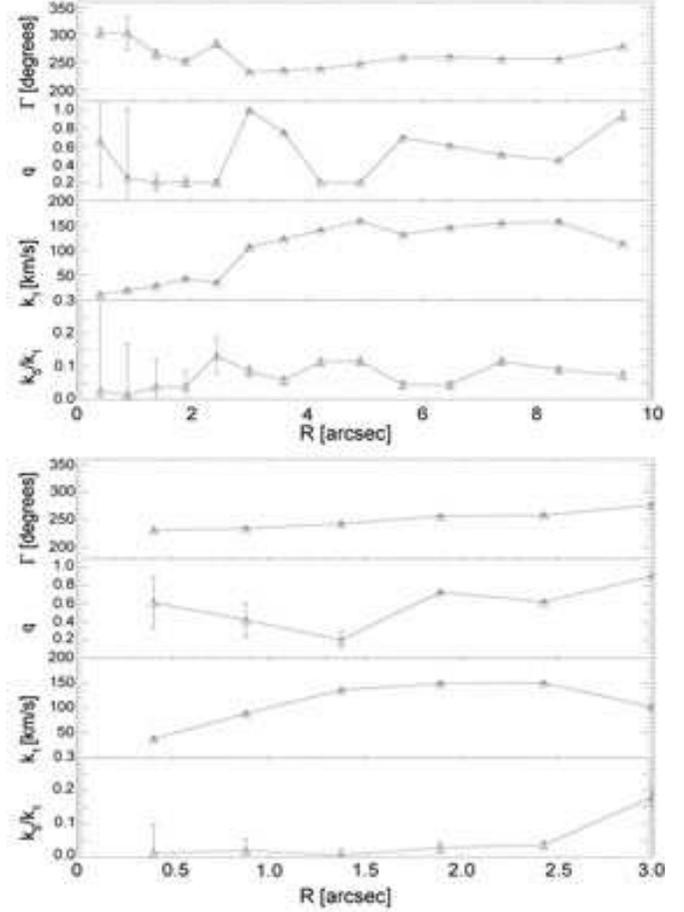}} \caption{Radial
profiles of the kinemetric properties, calculated using the
kinemetry programme for the interacting galaxy A of Fig.
\ref{disturbed0105} (minor merger) at redshift $z=0.1$ (top) and
$z=0.5$ (bottom). The position angle $\Gamma$ and the flattening $q$
of the best fitting ellipses as well as the first and the fifth
order Fourier terms $k_1$ and $k_5$ are plotted as a function of
radius.} \label{plotsdis}
\end{center}
\end{figure}

In Sect. \ref{extract} we have defined the kinematical class
'disturbed rotation' as a rotating system with a clearly visible
distortion. As an example for this class we investigate a model
galaxy undergoing a minor merger (simulation 2). At this snapshot
the distance between the two galaxies is smaller than the radius of
galaxy A, i.e. galaxy B encounters the Milky Way type galaxy A in
its first passage. The small galaxy permeating the more massive
galaxy A is clearly visible in the velocity field, as irregular
structure in the lower right part of the VF presented in Fig.
\ref{disturbed0105}. At redshift $z=0.1$ the kinemetry analysis
shows strong variations of the position angle $\Gamma$ and the
flattening $q$ with radius (see Fig. \ref{plotsdis}). The radial
fluctuations are stronger than for the isolated galaxy. The first
order Fourier term $k_1$, which corresponds to the rotation curve
also shows distortions and the ratio $k_5/k_1$ is also significantly
higher than in the isolated case. Note that the shape and the
maximum value of the RC are completely different from the regular
RC. Therefore, this galaxy could not be taken for Tully-Fisher
studies. We note, that also RCs extracted from this galaxy with a
slit show clear signatures of a distortion (Kronberger et al.,
2006), especially an asymmetric shape with a rising side pointing
towards the interaction. However, with the help of 2D velocity
fields the nature of the interaction becomes a lot more accessible.
The almost undistorted left hand side of the VF points towards a
very asymmetric interaction, which suggests a tidal interaction.
This asymmetry is less visible at $z=0.5$, where the distortion is
smeared out. Simple eye-balling might even lead to a
misclassification as 'undisturbed rotation'. However, by a more
detailed comparison with Fig. \ref{regular0105} some differences are
clearly visible. The most striking difference is the presence of a
clear kinematic axis in the regular VF, which is not present in Fig.
\ref{disturbed0105}.

\begin{figure}
\begin{center}
{\includegraphics[width=\columnwidth]{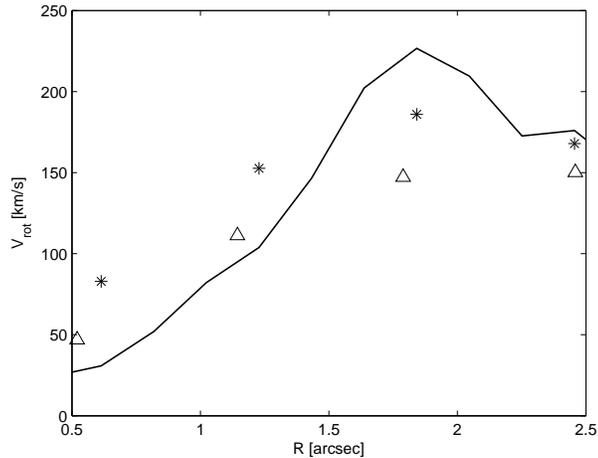}} \caption{Rotation
curves of the interacting galaxy extracted from the 2D velocity
field (triangles) and using a modelled slit (stars) at redshift
z=0.5. The solid line shows the corresponding RC from the 2D field
at redshift z=0.1.} \label{rcsdistorted}
\end{center}
\end{figure}

We present again the rotation curve extracted from the 2D velocity
field and using a simulated slit in Fig. \ref{rcsdistorted}. In
contrast to the case of the isolated galaxy, the RCs differ
significantly in the interacting system. The reason for this
difference is the fact that in an interacting system the kinematic
axis is not well defined and does in general not coincide with the
photometric axis of the system. Thus, the slit does in general not
follow the kinematic axis of the galaxy. Only the knowledge of the
full 2D velocity field allows to determine the kinematic axis.

The plots of the radial behaviour of the kinemetry parameters in
Fig. \ref{plotsdis} do not yield clear evidence for a distorted VF
at z=0.5. Only the outermost data point of $k_5/k_1$ shows a
significant increase. The situation is even worse for the resolution
obtained with FLAMES/GIRAFFE. Although the maximum rotational
velocity of the RC becomes similar to the one of the isolated galaxy
due to the smearing out of the VF, a misclassification as
undistorted galaxy would result in systematic errors in Tully-Fisher
studies. The luminosity of the system is possibly enhanced due to a
merger induced starburst.

\begin{figure}
\begin{center}
{\includegraphics[width=\columnwidth]{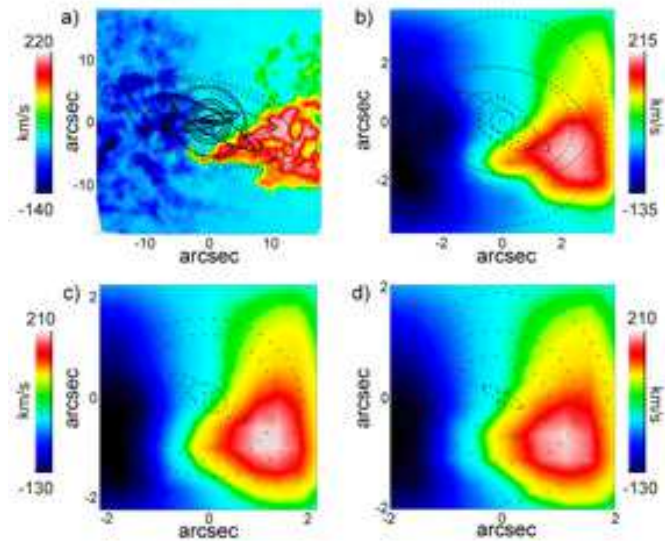}} \caption{Evolution
of the 2D velocity field appearance with redshift: a) $z=0.05$, b)
$z=0.3$ c) $z=0.8$ d) $z=1.0$. Overlayed as contours are the best
fitting ellipses from the kinemetry analysis. $z=0.1$ and $z=0.5$
are shown in Fig. \ref{disturbed0105}} \label{distevo}
\end{center}
\end{figure}

In Fig. \ref{distevo} we show the 2D velocity field of this example
for redshift $z=0.05$, $z=0.3$, $z=0.8$, and $z=1.0$. It is clearly
visible, how substructures are smeared out at higher redshift.

We studied the effects of different interaction geometries on the
results presented above. For this purpose we considered a
counter-rotating unequal mass merger (simulation 4) and an unequal
mass merger, where the disc of galaxy B is initially perpendicular
to the disc of galaxy A (simulation 5). The appearance of the
velocity fields does of course differ for the different interaction
scenarios. However, the main conclusions drawn so far do not change.
The merger is clearly visible in the asymmetric shape of the
velocity field, the distortions are, however, smeared out at higher
redshift.

\begin{figure}
\begin{center}
{\includegraphics[width=\columnwidth]{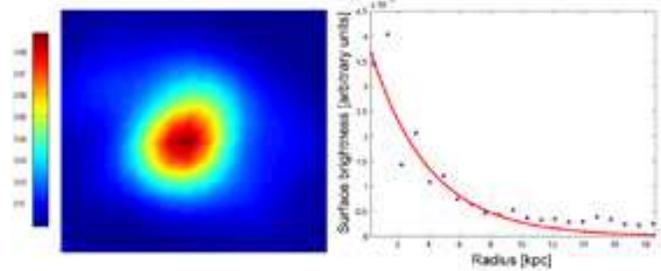}} \caption{The
distribution of the stellar mass in the galaxy as 2D image (left)
and as a radial profile (right), which translates under an
assumption of a constant mass-to-light ratio to a surface brightness
profile. The galaxy is shown 100 Myr after the first pericentre
passage of galaxy B at a resolution corresponding to an artificial
redshift of $z=0.5$. The exponential fit to the profile is shown as
a red line. The physical field of view of the 2D image is 36
$\times$ 36 kpc.} \label{direct_70}
\end{center}
\end{figure}

\begin{figure}
\begin{center}
{\includegraphics[width=\columnwidth]{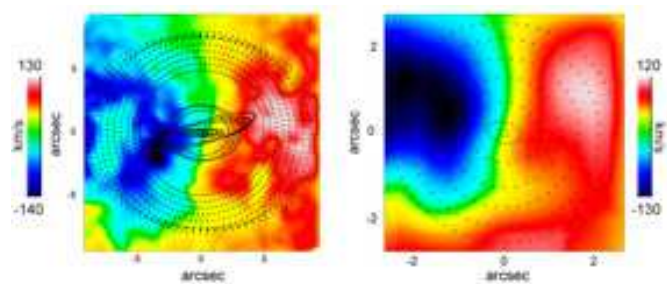}} \caption{2D
velocity field of a modelled disc galaxy 100 Myr after a minor
merger event as seen at redshift 0.1 (left) and 0.5 (right).
Overlayed as contours are the best fitting ellipses from the
kinemetry analysis.} \label{disturbed70}
\end{center}
\end{figure}

Of course it is more probable to observe an interacting system in a
pre- or post merger phase than in the relatively short period where
the two galaxies permeate each other. Therefore, we study the same
system 100 Myr after the direct encounter described above, i.e.
after the first pericentre passage. This is the period between the
first passage and the complete merging of the two galaxies. The
small galaxy went through the gaseous disc of the massive galaxy and
left a disturbed velocity field. In the direct image the interaction
is not visible anymore at redshift $z=0.5$. In Fig. \ref{direct_70}
we show the distribution of the stellar mass in the galaxy as 2D
image and as a radial profile. By assuming a certain constant
mass-to-light ratio this could be translated to a light
distribution. The profile can be fitted by an exponential law. The
velocity field shows still some signatures of the interaction (see
Fig \ref{disturbed70}) at lower redshift but at z=0.5 these are
heavily smeared out. Thus, the velocity information is an important
complement to the morphological analysis when studying interactions
of galaxies but only if the VF is sufficiently sampled.

Kinematic disturbances from interactions are expected to fade within
a few rotation cycles ($\leq$ 1 Gyr) (e.g. Dale et al. 2001,
Kronberger et al. 2006). Only 200 Myr after the direct encounter
presented above neither the direct image nor the VF show clear
signatures of the interaction anymore. In a more detailed analysis,
however, signs of the interaction can still be found. About 1 Gyr
after the first passage, when both galaxies are well separated but
approaching each other again, galaxy A would be classified as
undisturbed, even when observed at low redshift.

The analysis so far was made for the rather large galaxy A, which
has a radial scale length of about 4.5 kpc. Therefore, the velocity
field is sampled by many pixels. The situation is completely
different for smaller galaxies, as e.g. model galaxy B. We
investigate a 'disturbed rotation' of galaxy B using a snapshot of
simulation 3, an equal mass merger of two small galaxies B. Equal
mass mergers massively disturb the velocity field of the gas of the
interacting galaxies. Most of the gas is converted into stars by a
merger induced starburst or lost to the intergalactic medium by
tidal forces (e.g. Kapferer et al., 2005). We choose a snapshot
about 300 Myr after the first encounter, when again some regular
rotation has established.

\begin{figure}
\begin{center}
{\includegraphics[width=\columnwidth]{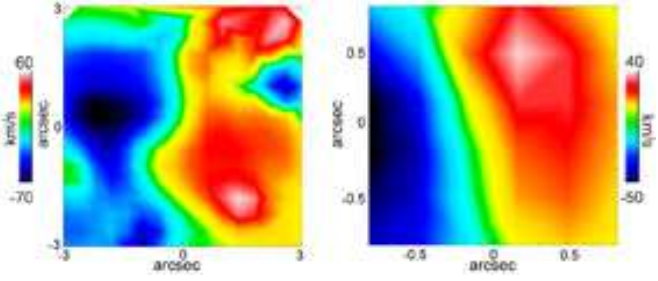}} \caption{2D
velocity field of a model galaxy B disturbed by an equal mass
merger, 300 Myr after the first pericentre passage of the two
galaxies, as seen at redshift 0.1 (left) and 0.5 (right).}
\label{equalvf}
\end{center}
\end{figure}

\begin{figure}
\begin{center}
{\includegraphics[width=\columnwidth]{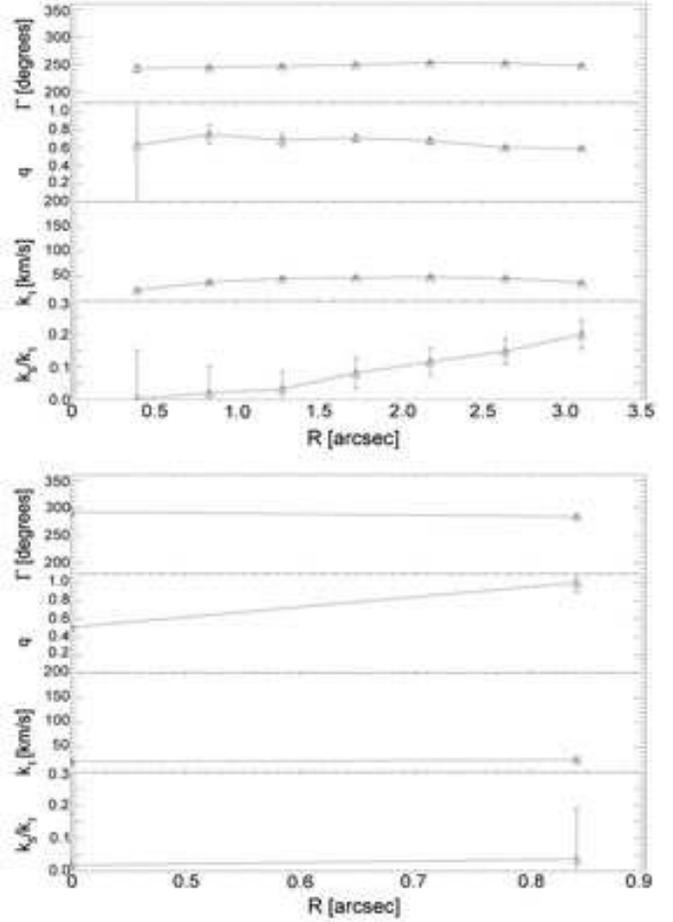}} \caption{Radial
profiles of the kinemetric properties, calculated using the
kinemetry programme for galaxy B 300 Myr after an equal mass merger
for redshift $z=0.1$ (top) and $z=0.5$ (bottom). The position angle
$\Gamma$ and the flattening $q$ of the best fitting ellipses as well
as the first and the fifth order Fourier terms $k_1$ and $k_5$ are
plotted as a function of radius.} \label{plotseq}
\end{center}
\end{figure}

In Fig. \ref{equalvf} the 2D velocity field of the galaxy for this
snapshot is shown at redshift $z=0.1$ and at $z=0.5$ with an angular
resolution of 0.4". While at $z=0.1$ the distortions in the VF are
clearly visible, they are completely smeared out at $z=0.5$ leaving
a regular VF, that would be classified as 'undisturbed rotation'.
This classification is supported by the radial behaviour of the
kinemetric quantities shown in Fig. \ref{plotseq}, although at
$z=0.5$ only two ellipses were fitted due to the small number of
available pixels. Such undetected distortions introduce an enormous
source of systematic errors to distant Tully-Fisher studies.
Therefore we conclude that current distant Tully-Fisher studies
cannot give reliable results for low-mass systems if the velocity
field is not sampled sufficiently.

\begin{figure}
\begin{center}
{\includegraphics[width=\columnwidth]{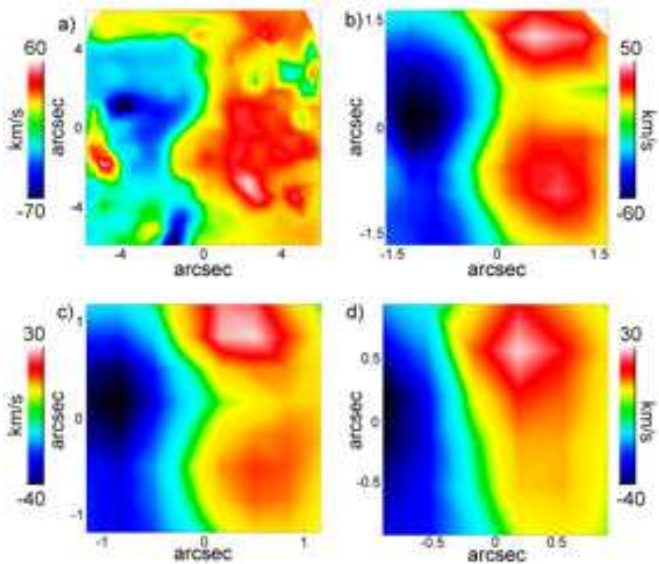}} \caption{Evolution
of the 2D velocity field appearance with redshift of a model galaxy
B disturbed by an equal mass merger, 300 Myr after the encounter: a)
$z=0.05$, b) $z=0.2$ c) $z=0.3$ d) $z=0.4$. Additionally the same 2D
velocity fields as seen at $z=0.1$ and $z=0.5$ are shown in Fig.
\ref{equalvf}} \label{distevo2}
\end{center}
\end{figure}

In Fig. \ref{distevo2} we present the 2D velocity field of galaxy B
for redshift $z=0.05$, $z=0.2$, $z=0.3$, and $z=0.4$. Gradually the
distortions are smeared out, leaving an almost regular VF at
$z=0.4$. Also for this interaction scenario we investigate how long
the 2D velocity field maintains the disturbed features. After the
first pericentre passage of the two galaxies the velocity field
settles again to an undisturbed state. As for the unequal mass
merger discussed above, the strongest features disappear after
several hundred Myr while an undisturbed velocity field is again
present after about 1 Gyr. Note that in this period the two galaxies
are approaching each other again for their second passage, but still
without any direct interaction yet.

\begin{figure}
\begin{center}
{\includegraphics[width=\columnwidth]{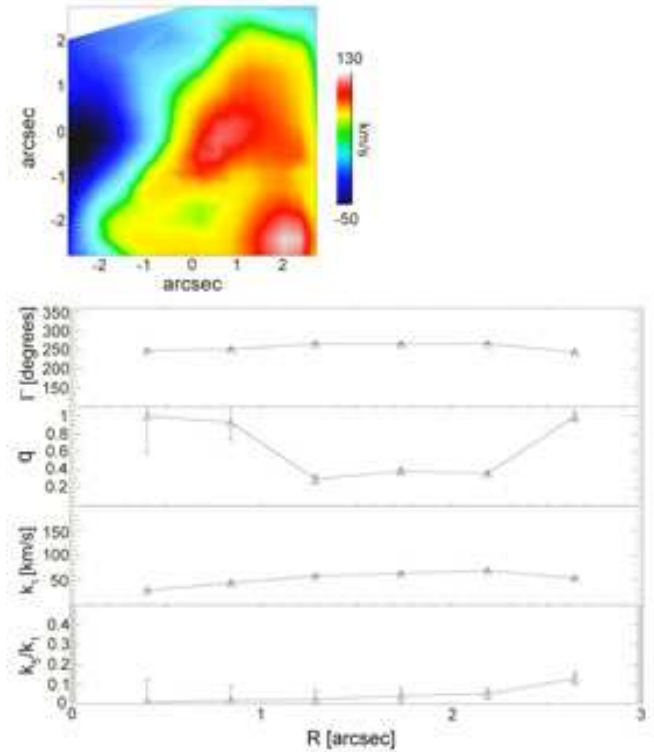}} \caption{The 2D
velocity field of a model galaxy A disturbed by an equal mass
merger, 100 Myr after the first passage of the two discs is shown in
the upper panel. The lower panel shows the radial profiles of the
kinemetric properties, calculated using the kinemetry programme. The
position angle $\Gamma$ and the flattening $q$ of the best fitting
ellipses as well as the first and the fifth order Fourier terms
$k_1$ and $k_5$ are plotted as a function of radius. Both figures
correspond to redshift z=0.5.} \label{equalAA}
\end{center}
\end{figure}

We also investigate a major merger between two large galaxies A
where the velocity field spreads over more pixels. In this case the
situation is, as expected, slightly better than for the small galaxy
B. Especially for the kinematical analysis more ellipses can be
fitted even at redshift z=0.5. Nevertheless a similar trend is
observable as for galaxy B. While the distortions are clearly
visible at low redshift, they are subsequently smeared out when the
galaxy is placed at higher redshift. In Fig. \ref{equalAA} we show
the 2D velocity field and the radial profiles of the kinemetric
properties as seen at redshift z=0.5 for model galaxy A disturbed by
an equal mass merger, 100 Myr after the first passage of the two
discs. The appearance of the velocity field and the radial behaviour
of mainly the flattening $q$ point towards an interaction (compare
to Figs. \ref{regular0105} and \ref{plotsreg} for the undisturbed
case). Small scale distortions are, however, smeared out.

\begin{figure}
\begin{center}
{\includegraphics[width=\columnwidth]{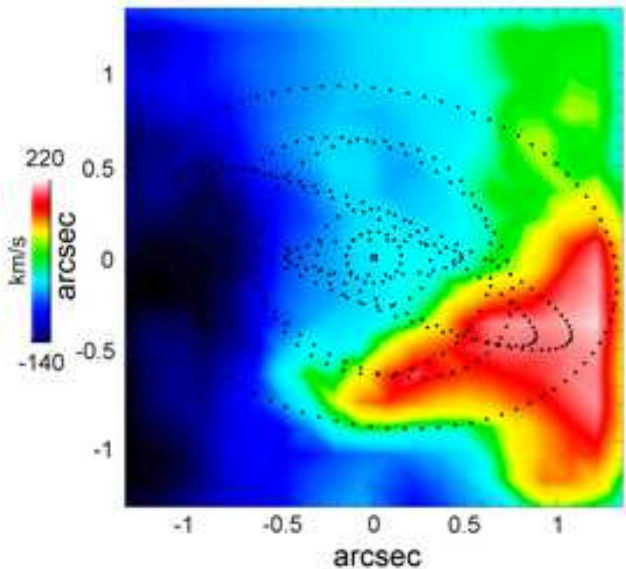}} \caption{2D
velocity field of a modelled disc galaxy disturbed by an ongoing
minor merger event as seen at redshift 2 using SINFONI at VLT.
Overlayed as contours are the best fitting ellipses from the
kinemetry analysis.} \label{sinfoni}
\end{center}
\end{figure}

Complementary to these investigations of VFs at intermediate
redshift, we study the appearance of the 2D velocity field at z=2 as
observed with SINFONI at VLT. Due to the use of adaptive optics,
these observations are not seeing limited and have a high angular
resolution (we adopt 0.15" for our study). As a test case we take
galaxy A during an ongoing minor merger in simulation 2. The
velocity field of this snapshot for seeing limited observations and
various redshifts was shown in Fig. \ref{distevo}. The galaxy has a
radial exponential disc scale length of 4.5 kpc and is therefore
comparable in size to the one observed by Genzel et al. (2006). The
velocity field of this snapshot as seen at redshift z=2 with an
instrument such as SINFONI is presented in Fig. \ref{sinfoni}.
Peculiarities in the VF caused by the merger are clearly visible.

\subsection{Distorted velocity field}

To heavily disturb the velocity field of a spiral galaxy, a tidal
interaction with a galaxy of similar or higher mass is necessary. In
this section we use the velocity field of the small galaxy B in
simulation 2 after its first passage through galaxy A, i.e. about
200 Myr after the first encounter.

\begin{figure}
\begin{center}
{\includegraphics[width=\columnwidth]{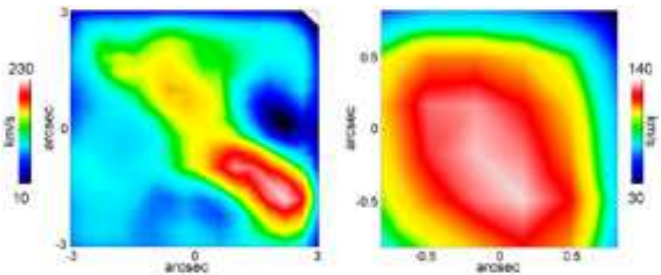}} \caption{2D
velocity field of a model galaxy B heavily disturbed by a merger
with galaxy A (200 Myr after the first passage through the disc of
galaxy A) shown at redshift 0.1 (left) and 0.5 (right).}
\label{distortvf}
\end{center}
\end{figure}

\begin{figure}
\begin{center}
{\includegraphics[width=\columnwidth]{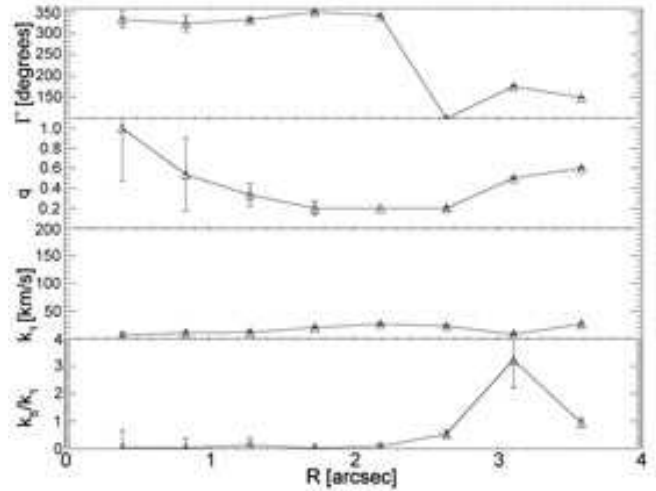}} \caption{Radial
profiles of the kinemetric properties, calculated using the
kinemetry programme for galaxy B 200 Myr after a major merger with
galaxy A for redshift $z=0.1$. The position angle $\Gamma$ and the
flattening $q$ of the best fitting ellipses as well as the first and
the fifth order Fourier terms $k_1$ and $k_5$ are plotted as a
function of radius.} \label{distortplot}
\end{center}
\end{figure}

In this case the VF shows no regular pattern of rotation anymore.
Both, at low and intermediate redshift the irregularity is visible,
although at $z=0.5$ the substructure is completely smeared out
leaving a single 'velocity blob' (see Fig. \ref{distortvf}). The bad
resolution of only 5 $\times$ 5 pixels at this redshift together
with the seeing is the reason for that. Also the better resolution
of 0.25" does not improve this image significantly. Obviously, such
a galaxy would not be used for a Tully-Fisher study and could not be
mistaken as undisturbed. Also the plots of the radial behaviour of
the kinemetry parameters clearly reflect the distortions in the VF.
Due to the low number of pixels at intermediate redshifts we
constructed this plot only for low redshifts. Fig. \ref{distortplot}
shows the kinematic parameters at $z=0.1$. A significant change of
the position angle $\Gamma$ coincides with a huge peak in the ratio
$k_5/k_1$ of about 3. Such a correlation between the position angle
and $k_5/k_1$ was also reported and discussed by Krajnovi\'{c} et
al. (2006).

Also for this interaction scenario the question arises for how long
the 2D velocity field maintains the severely distorted shape. We
found that in this case the distortions of the velocity field are so
strong, that the galaxy at no point between the first passage and
the final merging could be classified as 'disturbed' or even
'regular' anymore. Before the first pericentre passage the galaxy
features an undisturbed rotation, which is increasingly disturbed
when the tidal forces of galaxy A become stronger. Once galaxy B
passes the Milky Way type galaxy the velocity field is heavily
distorted, as discussed above.

\subsection{Discussion of observational biases}\label{biases}

Throughout the paper we have assumed that the velocity field of the
galaxy is always observable out to four disc scale lengths, which in
real observations is of course due to limited sensitivity not always
possible. Therefore, depending on the instrument, one might just
observe the inner parts of the VF. As a consequence distortions in
the outskirts can be missed. Here we were interested in the
principle effects of limited resolution on the appearance of the 2D
velocity fields, without specifying a special instrument, i.e. for
an idealized observation. In a future work we want to study
flux-limited velocity fields for special instruments for comparison
with observed data.

\begin{figure}
\begin{center}
{\includegraphics[width=\columnwidth]{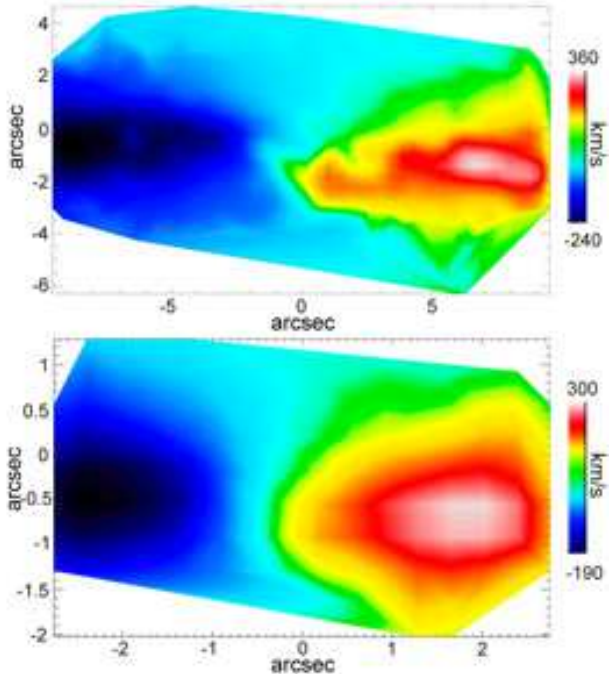}} \caption{2D
velocity field of a modelled disc galaxy disturbed by an ongoing
minor merger event at redshift 0.1 (top) and 0.5 (bottom). In
contrast to Fig. \ref{disturbed0105} the inclination in this figure
is i$=$80$^{\circ}$.} \label{inclination}
\end{center}
\end{figure}

A second parameter that changes the appearance of the 2D velocity
field is the inclination of the galaxy. Throughout the paper we have
adopted an inclination i$=$35$^{\circ}$. For an undisturbed galaxy
typically a correction with the sine of the inclination is applied
to the rotation curve. The appearance of a regular 2D field is not
severely affected by inclination. In the case of an interacting
galaxy, however, the appearance of the velocity field changes with
inclination. A systematic investigation of the effects of
inclination is, however, difficult, as they depend on the specific
interaction geometry. We sketch here the influence of inclination on
the VF presented in Sect. \ref{DR}. Instead of i$=$35$^{\circ}$, we
now choose i$=$80$^{\circ}$, i.e. nearly edge-on. The unequal mass
edge-on merger of simulation 2 produces VF distortions mainly in the
plane of the disc. These distortions are therefore well visible at
low inclinations, see Fig. \ref{disturbed0105} but get less
prominent for higher inclinations. In Fig. \ref{inclination} we show
the same snapshot as in Fig. \ref{disturbed0105} but at an
inclination of i$=$80$^{\circ}$ instead of i$=$35$^{\circ}$. At
$z=0.5$ the VF even appears regular, hence the interaction might not
be recognized.

\subsection{Implications for distant Tully-Fisher studies}

Many recent studies have analysed the luminosity evolution of
galaxies via the Tully-Fisher relation (e.g. Ziegler et al. 2003,
B\"ohm et al. 2004, Bamford et al. 2005, Nakamura et al. 2006,
Flores et al. 2006). Some of these studies also focussed on
environmental effects, i.e. differences in the Tully-Fisher relation
between field and cluster population. Ziegler et al. (2003) and
Nakamura et al. (2006) find no significant differences between field
and cluster galaxies, whereas Bamford et al. (2005) claim that
galaxies in cluster are on average brighter than their field
counterparts. The Tully-Fisher relation (TFR) is also extensively
used as test-bench for galaxy formation models. Portinari \&
Sommer-Larsen (2007) investigated the redshift evolution of the
Tully-Fisher relation in cosmological simulations. They find an
offset between the observed and simulated Tully-Fisher relation at
z=0. The evolution they find is intermediate between diverse
observational results. Parts of the discrepancies between the
various observational results may be attributed to the way
distortions in the velocity fields of the galaxies are accounted
for.

The results from this work suggest that galaxies from an observed
sample which are misclassified as undisturbed can introduce a
systematic error on the Tully-Fisher diagram. A Tully-Fisher
analysis assumes that the galaxies are in virial equilibrium, which
is revealed by a smooth, regular and symmetric rotation curve, that
rises rapidly in the inner part and turns over into a flat part.
This "flat" or maximum circular velocity is one parameter of the
TFR, the other being the luminosity. In a disturbed system, the
measured V$_{max}$ is not an accurate estimate for the virial
velocity of the halo anymore and, hence, no good proxy for the total
mass of the system. Also, the amount of luminosity of the galaxy is
influenced by the interaction (see Kapferer et al. 2005). To which
extent both quantities are affected by the interaction depends on
the interaction geometry and the time.

Therefore, if distortions or peculiarities are visible in the RC,
the galaxy would not be included in a TF analysis. However in this
study, we have shown that at intermediate redshifts due to lack of
resolution, signatures of an interaction are smeared out, so that a
distorted galaxy mimics a regular one in several cases. If such
disturbed (but unrecognized) systems are used or a Tully-Fisher
study, the scatter will be significantly increased. In contrast to
slit observations of a one-dimensional RC, the availability of a
two-dimensional VF allows a much better assessment of irregularity.
Nevertheless, in case of a small and faint galaxy, where the VF
spreads over a few pixels only, an interaction could be missed as
well. The inclusion of such intrinsically disturbed objects could
therefore mimic an evolution at the faint end of the Tully-Fisher
relation.

\begin{figure}
\begin{center}
{\includegraphics[width=\columnwidth]{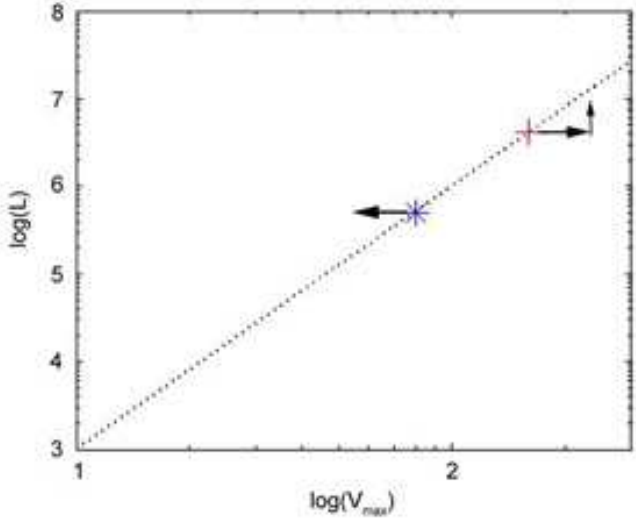}} \caption{Position
of the two undisturbed model galaxies in the luminosity-rotational
velocity plane (red cross: galaxy A; blue asterisk: galaxy B). The
two galaxies lie on a Tully-Fisher relation with slope 3 (dotted
line), consistent with observations of local galaxy samples. The
arrows indicate the trends in which way the interactions studied in
this work change the positions of the galaxies. For the Milky Way
type galaxy, the investigated interaction is the minor merger with
galaxy B, while for model galaxy B it is the major merger case with
a second galaxy B. These trends can be observed at all redshifts but
in a different strength.}\label{tf}
\end{center}
\end{figure}

We show in Fig. \ref{tf} how a galaxy moves in the Tully-Fisher
diagram if V$_{max}$ values are used that do not correspond to the
virial velocity. First, we place the undisturbed galaxies onto the
TFR under the assumption of a constant mass-to-light ratio. The
simulated Milky Way galaxy A and the less massive galaxy B then
follow a linear relation with slope of 3, i.e. L $\propto$
V$_{max}^3$. This is in good agreement with observed local values.
Pierce \& Tully (1992) get a value of 2.99 for their sample, other
authors get values between 2.96 and 3.2 (see Ziegler et al. 2002 and
references therein). In Fig. \ref{tf} we then sketch how
interactions change the positions of the galaxies. Note that these
trends are taken from the interaction scenarios studied in this
work. Other interaction geometries or times might lead to a
different effect. The Milky Way type galaxy A moves to the right, if
the minor merger presented in Sect. \ref{DR} is not recognized. As
the interaction also triggers star formation, the luminosity
increases at the same time. In the case studied here, the effect on
V$_{max}$ is greater than the brightening of the galaxy. For the
small galaxy B in the case of a major merger the interactions lead
to an underestimation of the true V$_{max}$. At the same time, the
luminosity stays roughly constant or even decreases by a small
amount, as stars are lost by the merger. Assuming this trend holds
more generally and that distortions are more easily missed at higher
redshift for small galaxies, this effect could mimic a luminosity
evolution at the faint end of the Tully-Fisher relation. Therefore,
a careful analysis is necessary to prevent the inclusion of
disturbed systems in distant Tully-Fisher studies.

\section{Summary and conclusions}

We have investigated 2D velocity fields of isolated and interacting
spiral galaxies using N-body/SPH simulations. We focussed on the
question how the full 2D velocity field of a galaxy can be used to
gain information on its internal kinematics. This issue was analysed
with special emphasis on distant Tully-Fisher studies. To summarise
and conclude:

\bit
\item We found that with the help of 2D velocity fields the nature
of the interaction becomes a lot more accessible than with rotation
curves from simple long-slit spectroscopy. Quantitative analysis
such as an harmonic expansion, which is, for example, used by the
kinemetry package of Krajnovi\'{c} et al. (2006), offer an
additional possibility to identify distortions in a VF.

\item Tidal interactions lead to an asymmetric velocity field, where
the side pointing at the interaction is disturbed while the side
remote from the interaction stays relatively unaffected. This
behaviour can also be found in the rotation curve of the system
(Kronberger et al. 2006).

\item By shifting the VF artificially to higher redshifts we found
that, although small scale structures in the VF are smeared out,
distortions are still visible at intermediate redshifts for large
galaxies. In the case of small galaxies even strong distortions are
not visible in the velocity field at $z\approx0.5$ with currently
available angular resolution.

\item Severely distorted kinematics, i.e. velocity fields with no or
just a small rotational component are identifiable also for small
galaxies at intermediate redshift. Thus they should not be mistaken
as undisturbed.

\item If the galaxy is undisturbed the quantities derived from the
velocity field, as e.g. the maximum rotational velocity V$_{max}$,
do not show systematic variations with resolution and are therefore
useable for Tully-Fisher studies.

\item We showed that with the help of adaptive optics near-infrared
spectrographs (e.g. SINFONI) it is possible to study the internal
kinematics of galaxies even at high-redshift (z$\sim$2), given that
the observed flux is sufficient to construct a 2D velocity field.
Ongoing merger events are then clearly visible in the VF.

\item Disturbed velocity fields settle again to a relaxed state
after about 1 Gyr. However, already after several hundred Myr the
distortions can get damped such that a misclassification as
'undisturbed rotation' is possible at higher redshifts.

\item With the assumption of a constant mass-to-light ratio, we
showed that the undisturbed model galaxies lie on a Tully-Fisher
relation with a slope of $~$3, which is consistent with observations
of local galaxy samples. Using systems with unrecognized distortions
for a Tully-Fisher study will significantly increase the scatter in
the observed relation. As it is more likely, that an interaction is
missed for small, faint galaxies, the inclusion of disturbed systems
can mimic an evolution in the faint end of the Tully-Fisher
relation.

\eit

\section*{Acknowledgements}

We thank the anonymous referee for fruitful comments which improved
the quality of the paper. The authors would like to thank Volker
Springel for providing them with GADGET2 and his initial-conditions
generator. We are also thankful to Davor Krajnovi\'{c} for his
Kinemetry software. Thomas Kronberger is a recipient of a DOC
fellowship of the Austrian Academy of Sciences. The authors further
acknowledge the UniInfrastrukturprogramm des BMWF Forschungsprojekt
Konsortium Hochleistungsrechnen, the German Science Foundation (DFG)
through Grant number Zi 663/6-1, and the Volkswagen Foundation (I/76
520). In addition, the authors acknowledge the ESO
Mobilit\"atsstipendien des BMWF (Austria), and the Tiroler
Wissenschaftsfonds (Gef\"ordert aus Mitteln des vom Land Tirol
eingerichteten Wissenschaftsfonds). We thank Elif Kutdemir for
fruitful discussion.

\end{document}